# A Two Dimensional Tunneling Resistance Transmission Line Model for Nanoscale Parallel Electrical Contacts

Sneha Banerjee[1], John Luginsland[1,2], and Peng Zhang[1]*

[1]*Department of Electrical and Computer Engineering, Michigan State University, East Lansing, Michigan 48824-1226, USA*

[2]*Present address: Confluent Sciences, LLC*

*email: pz@egr.msu.edu

Contact resistance and current crowding are important to nanoscale electrical contacts. In this paper, we present a self-consistent model to characterize partially overlapped parallel contacts with varying specific contact resistivity along the contact length. For parallel tunneling contacts formed between contacting members separated by a thin insulating gap, we examine the local voltage-dependent variation of potential barrier height and tunneling current along the contact length, by solving the lumped circuit transmission line model (TLM) equations coupled with the tunneling current self consistently. The current and voltage distribution along the parallel tunneling contacts and their overall contact resistance are analyzed in detail, for various input voltage, electrical contact dimension, and material properties (i.e. work function, sheet resistance of the contact members, and permittivity of the insulating layer). It is found the existing one-dimensional (1D) tunneling junction models become less reliable when the tunneling layer thickness becomes smaller or the applied voltage becomes larger. In these regimes, the proposed self-consistent model may provide a more accurate evaluation of the parallel tunneling contacts. For the special case of constant ohmic specific contact resistivity along the contact length, our theory has been spot-checked with finite element method (FEM) based numerical simulations. This work provides insights on the design, and potential engineering, of nanoscale electrical contacts with controlled current distribution and contact resistance via engineered spatially varying contact layer properties and geometry.

## I. INTRODUCTION

Nanoscale electrical contacts have attracted substantial attention due to the advancements in nanotechnology, material sciences and growing demands for miniaturization of electronic devices and high packing density. Contact resistance and their electro-thermal effects have become one of the most critical concerns of very large scale integration (VLSI) circuit designers, because of the excessive amount of Joule heating being deposited at the contact region [1–6]. The electrical contact properties have been extensively studied in metal-semiconductor [7–9], metal-insulator-semiconductor and metal-insulator-metal [10–13] junctions. The growing popularity of novel electronic circuits based on graphene, carbon nanotubes (CNTs) and other new materials has made contact engineering crucial. CNT based devices, in particular, experience significant challenges because of the inter-tube connections. On macroscopic level, the exceptional intrinsic electrical properties [14,15] of CNTs become elusive [3,14,16]. Contact resistances between CNTs profoundly affect the electron transport and reduce the electrical conductivity of carbon nanofiber (CNF) [15–

[17], and greatly limit the performance of CNT thin film based Field Effect Transistors (FETs) [18–21]. One can naturally expect these issues also arising from other novel two-dimensional materials (boron nitride, molybedenum sulfide, black phosphorus, etc) as well as new nano-composites. While the work presented here is generalizable to these other material systems, here we choose carbon materials as examples.

Tunneling type of electrical contacts[11,22–25] are commonly found for CNT-CNT [16,22,26–30], CNT-Metal [31–33] and CNT-graphene [34,35] contacts, where the contacting members are separated by very thin insulating layers. Tunneling effects in contact junctions significantly lower the electrical conductivity of the CNT/polymer composite thin films [23]. It is also found that tunneling resistance plays a dominant role in the electrical conductivity of CNT-based polymeric or ceramic composites [27].

For decades, the basic models of tunneling current between electrodes separated by thin insulating films have been those of Simmons [36–38] in 1960s. Simmon's formula have since been used for evaluating tunneling current in tunneling junctions [24,29,39]. Though there have been attempts to extend Simmons' models to the field emission and space-charge-limited regimes [11,40,41], it is always assumed that the tunneling junctions are one-dimensional (1D), i.e. there is no variations on the voltages drops along the length of the tunneling junction and the insulating film thickness is uniform. Thus, these existing models of tunneling junctions give no hint on the variation of tunneling current along the contact length and the importance of current crowding near the contact area when the two contacting members are partially overlapping (cf. Fig. 1). On the other hand, the widely used transmission line models (TLM) for electrical contacts typically assume the contact resistivity of the interface layers are constant[10,42–44]. It is questionable to apply these models to study the tunneling contacts, as the tunneling resistance depends on the junction voltage that varies spatially along the contact length.

In this paper, we propose a two-dimensional (2D) transmission line model for partially overlapped parallel contacts with spatially varying specific contact resistivity. Spatial dependence of specific contact resistivity of the contact interface may be introduced by many factors, such as nonuniform distribution of the resistive contaminants, oxides, or foreign objects at the contact interface, formation of contact interfaces with spatially varying thickness, or the presence of tunneling contacts between contact members. In the latter case, because of the nonlinear current-voltage characteristics of the tunneling junctions[11,36], the specific resistivity along the contact length will become spatially dependent, even for a tunneling layer with uniform thickness (Fig. 1). For the tunneling-type contacts, the model considers the variation of potential barrier height and tunneling current along the contact length, by solving the TLM equations coupled with the tunneling current self consistently. We provide comprehensive analysis of the effects of contact geometry (i.e. dimension of the contact, and distance between the contact electrodes), and material properties (i.e. work function, sheet resistance of the contact members, and permittivity of the insulating layer) on the spatial distributions of currents and voltages across these contacts, and the overall contact resistance of parallel contacts.

In Sec. II, the formulation of our 2D contact resistance TLM model is given. We would like to point out that, albeit an application of the standard transmission line theory based on the Kirchhoff's laws, the TLM has been used extensively with great success to characterize mesoscale and nanoscale electrical contacts[10,30,41,43]. Here we further extend the TLM model with the effects of spatially dependent contact resistivity. Results and discussions are presented in Sec. III, where we have considered three cases of parallel contacts: 1) constant specific contact resistivity, 2) linearly varying specific contact resistivity, and 3) tunneling contact resistivity depending on local junction voltages along the contact length. The first case of uniform specific contact resistivity along the contact length has been verified with COMSOL [45] 2D simulations. For the last case, for simplicity, we use the Simmons' model [35-37] to determine the local current-voltage characteristics across the tunneling junction. Though full scale quantum mechanical calculations may have to be used to accurately evaluate the nanoscale circuits, our model based on Simmons formula reveals the fundamental scalings and parametric dependence of current and voltage profiles, as well as electric contact resistance of tunneling contacts. Summary and suggestions for future research are given in Section IV.

Note that, although this paper is focused on the normal Schrodinger tunneling type electrical contacts, the proposed TLM with spatially varying contact resistivity can be used for many other types of electric contacts, such as nanoscale Schottky contacts based on 2D materials heterostructure[46,47], and Klein tunneling junctions[48].

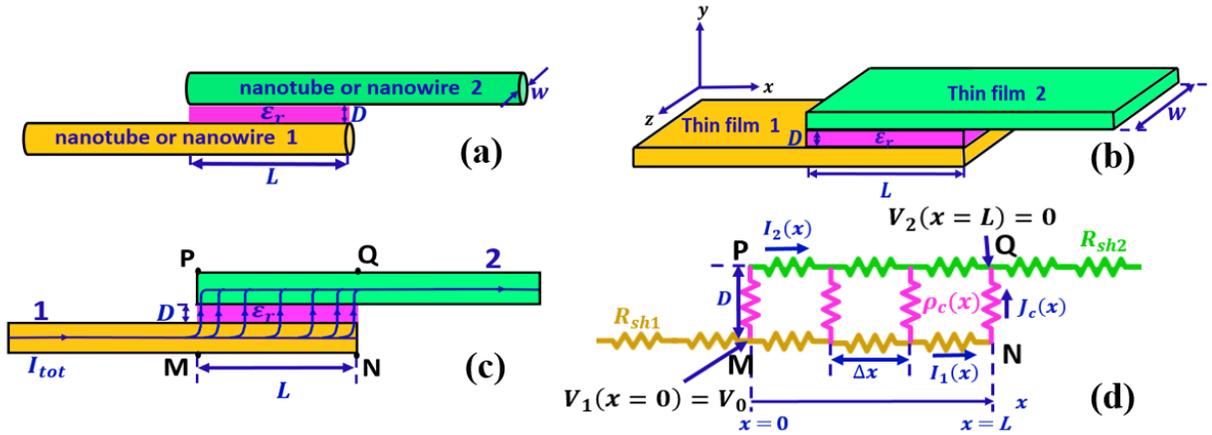

FIG. 1. A parallel, partially overlapped electric contact. The contacts are formed between (a) nanotube or nanowire 1 and 2, and (b) thin film 1 and 2; (c) side view of the contact; (d) its transmission line model. In (a), (b) and (c) a thin resistive interface layer (or a tunneling layer of permittivity $\varepsilon_r$) is sandwiched between the two contacting members.

## II. FORMULATION

Consider a parallel contact formed between two nanowires or nanotubes or between two conducting thin films or layers, as shown in Figs. 1(a) and 1(b), respectively. The distance between the two contact members is $D$, and the contact length is $L$. A thin resistive interface layer is sandwiched between them. Both contacts in Figs. 1(a) and 1(b) can be described by a two-dimensional (2D) model, as shown in Fig. 1(c). Note that the proposed formulation is generally applicable to parallel Cartesian nanojunctions with different shape of the electrodes, for example, electrical contact between a nanowire and a thin film. In the 2D model, the effects of the transverse dimension (perpendicular to the paper) can be included in the effective sheet resistances $R_{sh1}$ and $R_{sh2}$ for conductor 1 and 2, respectively, such that there is no variation along the width $w$ in the transverse dimension. The spatial dependent specific interfacial resistivity (also termed specific contact resistivity) is $\rho_c(x)$, which is either predefined, or calculated from the local tunneling current in case of insulating tunneling layer [36–38]. We use the DC equivalent lump circuit transmission line model (TLM) [10,42–44], as shown in Fig. 1(d), to model the 2D parallel contact in Fig. 1(c).

In the contact region PQNM in Figs. 1(c) and 1(d), using Kirchoff's laws for current and voltage, we get the following equations,

$$I_1(x) - I_1(x + \Delta x) = \frac{V_1(x) - V_2(x)}{\rho_c(x)} \Delta x\, w, \quad (1a)$$

$$V_1(x) - V_1(x + \Delta x) = I_1(x)\, R_{sh1} \Delta x / w, \quad (1b)$$

$$I_2(x + \Delta x) - I_2(x) = \frac{V_1(x) - V_2(x)}{\rho_c(x)} \Delta x\, w, \quad (1c)$$

$$V_2(x) - V_2(x + \Delta x) = I_2(x)\, R_{sh2} \Delta x / w, \quad (1d)$$

where $I_1(x)$ and $I_2(x)$ represent the current flowing at $x$ through the lower contact member, MN and upper contact member, PQ respectively, and $V_1(x)$ and $V_2(x)$ the local voltage at $x$ along MN and PQ, respectively, and $w$ is the effective transverse dimension of the contacts. When $\Delta x \to 0$, Equation (1) becomes,

$$\frac{\partial I_1(x)}{\partial x} = -w J_c(x), \quad (2a)$$

$$\frac{\partial V_1(x)}{\partial x} = -\frac{I_1(x) R_{sh1}}{w}, \quad (2b)$$

$$\frac{\partial I_2(x)}{\partial x} = w J_c(x), \quad (2c)$$

$$\frac{\partial V_2(x)}{\partial x} = -\frac{I_2(x) R_{sh2}}{w}, \quad (2d)$$

where $J_c(x) = V_g(x)/\rho_c(x)$ and $V_g(x) = V_1(x) - V_2(x)$ are the local current density and the local voltage drop across the contact interface at $x$, respectively.

Note that, from Eqs. (2a) and (2c), $I_1(x) + I_2(x) = I_{tot}$ = constant, where $I_{tot}$ is the total current in the circuit, to be determined from the boundary conditions. The boundary conditions for Eq. (2) are,

$$V_1(x = 0) = V_o, \quad (3a)$$
$$I_2(x = 0) = 0, \quad (3b)$$
$$I_1(x = L) = 0, \quad (3c)$$
$$V_2(x = L) = 0, \quad (3d)$$

where, without loss of generality, we assume the voltage of the upper contact member at $x = L$ is 0, and the externally applied voltage at $x = 0$ of the lower contact member is $V_o$. Note that $I_1(x = 0) = I_{tot}$, and $I_2(x = 0) = 0$. From Eqs. (2) and (3), it is easy to show $V_1'(x = 0) = -I_{tot} R_{sh1}/w$, $V_1'(x = L) = 0$, $V_2'(x = 0) = 0$, $V_2'(x = L) = -I_{tot} R_{sh2}/w$, where a prime denotes a derivative with respect to $x$. For the contact model in Fig. 1(d), the contact resistance is defined as,

$$R_c = \frac{V_1(0) - V_2(L)}{I_{tot}} = \frac{V_o}{I_{tot}}. \qquad (4).$$

It is convenient to introduce non-dimensional quantities, $\bar{x} = x/L$, $\bar{\rho}_c(\bar{x}) = \rho_c(x)/\rho_{c0}$, $\bar{R}_{sh2} = R_{sh2}/R_{sh1}$, $\bar{I}_1(\bar{x}) = I_1(x)/I_o$, $\bar{I}_2(\bar{x}) = I_2(x)/I_o$, $\bar{J}_c(\bar{x}) = J_c(x)LW/I_o$, $\bar{V}_1(\bar{x}) = V_1(x)/V_o$, $\bar{V}_2(\bar{x}) = V_2(x)/V_o$, $\bar{V}_g(\bar{x}) = V_g(x)/V_o$, and $\bar{R}_c = R_c/R_{c0}$, where we define $I_o = wV_0/R_{sh1}L$, $\rho_{c0} = V_0wL/I_o$, and $R_{c0} = R_{sh1}L/w$. In normalized forms, Eq. (2) can be recast into the following second order differential equations,

$$\frac{\partial^2 \bar{V}_1(\bar{x})}{\partial \bar{x}^2} = \bar{J}_c(\bar{x}), \qquad (5a)$$

$$\frac{\partial^2 \bar{V}_g(\bar{x})}{\partial \bar{x}^2} = (1 + \bar{R}_{sh2})\bar{J}_c(\bar{x}), \qquad (5b)$$

$$\bar{\rho}_c(\bar{x}) \frac{\partial^2 \bar{I}_1(\bar{x})}{\partial \bar{x}^2} + \frac{\partial \bar{\rho}_c(\bar{x})}{\partial \bar{x}} \frac{\partial \bar{I}_1(\bar{x})}{\partial \bar{x}} - (1 + \bar{R}_{sh2})\bar{I}_1(\bar{x}) + \alpha \bar{R}_{sh2} = 0, \qquad (5c)$$

where $\bar{J}_c(\bar{x}) = \bar{V}_g(\bar{x})/\bar{\rho}_c(\bar{x})$, and $\bar{V}_g(\bar{x}) = \bar{V}_1(\bar{x}) - \bar{V}_2(\bar{x})$. The corresponding boundary conditions to Eqs. 5(a)-5(c) are respectively,

$$\bar{V}_1(\bar{x} = 0) = 1, \bar{V}_1'(\bar{x} = 0) = -\alpha \text{ and } \bar{V}_1(\bar{x} = 1) = \bar{V}_g(\bar{x} = 1), \quad (6a)$$

$$\bar{V}_g'(\bar{x} = 0) = -\alpha, \quad \bar{V}_g'(\bar{x} = 1) = \alpha \bar{R}_{sh2}, \quad (6b)$$

$$\bar{I}_1(\bar{x} = 0) = \alpha, \quad \bar{I}_1(\bar{x} = 1) = 0, \quad (6c)$$

where the unknown constant $\alpha = I_{tot}/I_o$ is the normalized total current in the circuit, and prime denotes a derivative with respect to $\bar{x}$. Note that integrating Eq. (5b) subject to Eq. (6b) gives $\int_0^1 \bar{J}_c(\bar{x}) d\bar{x} = \alpha$, which means that the total current is conserved across the contact interface.

Equations (5) and (6) are solved to give the voltage and current distribution along and across the contact interface as well as the total contact resistance, for a given electrical contact (Fig. 1) with spatially dependent interface specific contact resistivity $\bar{\rho}_c(\bar{x})$. An example of the procedure to solve Eqs. (5) and (6) numerically is as follows. For an initially guess on $\alpha$, Eq (5b) is solved using the shooting method, subject to Eq. (6b). Next, Eq (5a) is solved with the initial values of $\bar{V}_1(0)$ and $\bar{V}_1'(0)$ from Eq. (6a). It is then checked whether $\bar{V}_1(1)$ is equal to $\bar{V}_g(1)$, as in Eq. (6a). The above-mentioned process repeats for different input $\alpha$ until the condition $\bar{V}_1(1) = \bar{V}_g(1)$ is satisfied. Finally, Eq (5c) is solved to get $\bar{I}_1$ (and $\bar{I}_2$).

In principle, Eqs. (5) and (6) can be solved numerically for arbitrary spatial dependence of specific contact resistivity $\bar{\rho}_c(\bar{x})$. Here, we focus on a few special cases of practical importance. We first consider the case of constant $\bar{\rho}_c$, where analytical solutions can be obtained (Sec. III Case 1), which also serve to validate our numerical approach. We then consider the effects of spatially dependent $\bar{\rho}_c(\bar{x})$ on the parallel electrical contacts. We focus on two situations: linearly varying specific contact resistivity along $x$ (Sec. III Case 2), and thin tunneling junction with uniform thickness (Sec. III Case 3), where analytical solutions to the TLM current and voltage equations are no longer available, and Eqs. (5) and (6) are solved numerically.

## III. RESULTS AND DISCUSSION

### Case 1: Constant specific contact resistivity $\rho_c$ along the contact length $L$

For the special case of constant specific contact resistivity $\rho_c$, the TLM equations, Eqs. (5) and (6), can be solved analytically to give,

$$\bar{I}_1(\bar{x}) = \frac{q}{K}[\sinh q(1-\bar{x}) + \bar{R}_{sh2}(\sinh q - \sinh q\bar{x})] \quad (7a)$$

$$\bar{I}_2(\bar{x}) = \frac{q}{K}[\sinh q(\bar{x}-1) + \bar{R}_{sh2}\sinh q\bar{x} + \sinh q] \quad (7b)$$

$$\bar{J}_c(\bar{x}) = \frac{q^2}{K}[\cosh q(1-\bar{x}) + \bar{R}_{sh2}\cosh q\bar{x}] \quad (7c)$$

$$\bar{V}_1(\bar{x}) = \frac{1}{K}[\cosh q(1-\bar{x}) + \bar{R}_{sh2}M + \bar{R}_{sh2}q(1-\bar{x})\sinh q] \quad (7d)$$

$$\bar{V}_2(\bar{x}) = \bar{V}_1(\bar{x}) - \bar{\rho}_c\bar{J}_c(\bar{x}) \quad (7e)$$

and $\quad \bar{R}_c = \frac{(1+\bar{R}_{sh2}^2)\cosh q + \bar{R}_{sh2}(2+q\sinh q)}{(1+\bar{R}_{sh2})q\sinh q} \quad (8)$

where $q = \frac{L}{\lambda_0} = \sqrt{\frac{1+\bar{R}_{sh2}}{\bar{\rho}_c}}$, $K = (1+\bar{R}_{sh2}^2)\cosh q + \bar{R}_{sh2}(2+q\sinh q)$ and $M = \cosh q\bar{x} + 1 + \bar{R}_{sh2}\cosh q$.

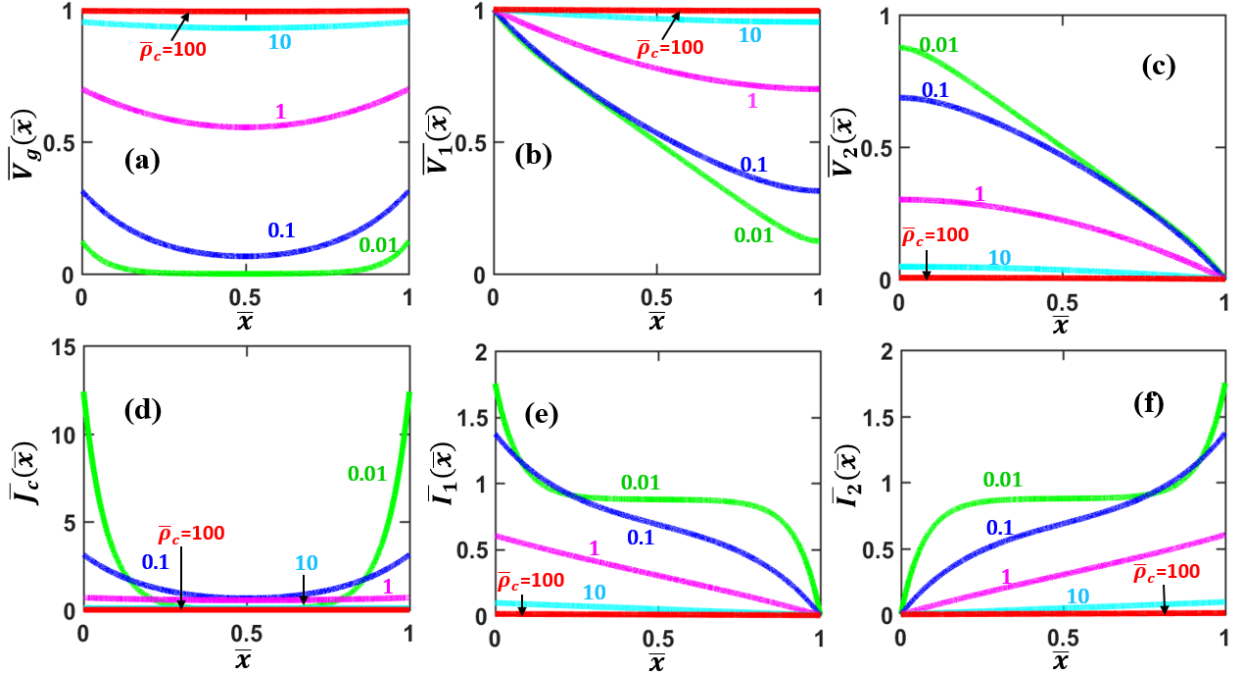

FIG. 2. Voltage and current profiles along similar parallel contacts with uniform contact resistivity. (a) Voltage drop across the contact interface $\bar{V}_g(\bar{x})$, voltage along (b) contact member 1 (MN), $\bar{V}_1(\bar{x})$, (c) contact member 2 (PQ), $\bar{V}_2(\bar{x})$, (d) current density across the contact interface $\bar{J}_c(\bar{x})$, current along (e) contact member 1, $\bar{I}_1(\bar{x})$, and (f) contact member 2, $\bar{I}_2(\bar{x})$, for different values of specific contact resistivity $\bar{\rho}_c$, for $\bar{R}_{sh2} = R_{sh2}/R_{sh1} = 1$. All the quantities are in their normalized forms defined in Sec. II.

Figure 2 shows the current and voltage distributions along the contact length and across the contact interface for various specific contact resistivity $\bar{\rho}_c$, for a parallel contact formed between similar contact members, $\bar{R}_{sh2} = R_{sh2}/R_{sh1} = 1$. The voltage along both contact members $\bar{V}_1$ and $\bar{V}_2$ decrease with $\bar{x}$, as shown in Figs. 2(b) and 2(c), respectively. The current $\bar{I}_1$ in contact member 1 decreases with $\bar{x}$ (Fig. 2(e)), whereas $\bar{I}_2$ in contact member 2 increases with $\bar{x}$ (Fig. 2(f)), with the total current $\bar{I}_1(\bar{x}) + \bar{I}_2(\bar{x})$ being kept a constant along $\bar{x}$. The profiles of both normalized voltage drop $\bar{V}_g(\bar{x})$ and current density $\bar{J}_c(\bar{x})$ across the interface layer, are symmetric along the contact length, with the minimum at the center of the contact structure $\bar{x} = 0.5$ and the maximum at the contact edges, as shown in Figs. 2(a) and 2(d), respectively. The current crowding effects near the contact edges are well-known phenomena [42,49], as the current density is distributed to follow the least resistive path (i.e. minimum overall resistance). It is important to note that as the specific contact resistivity $\bar{\rho}_c$ decreases, the interface current density $\bar{J}_c$ becomes more crowded towards the contact edges, as shown in Fig. 2(d). In other words, the less resistive the contact interface layer, the more severe of the current crowding effects, which is in agreement with previous studies using both TLM [42,44] and field theory [10,49,50].

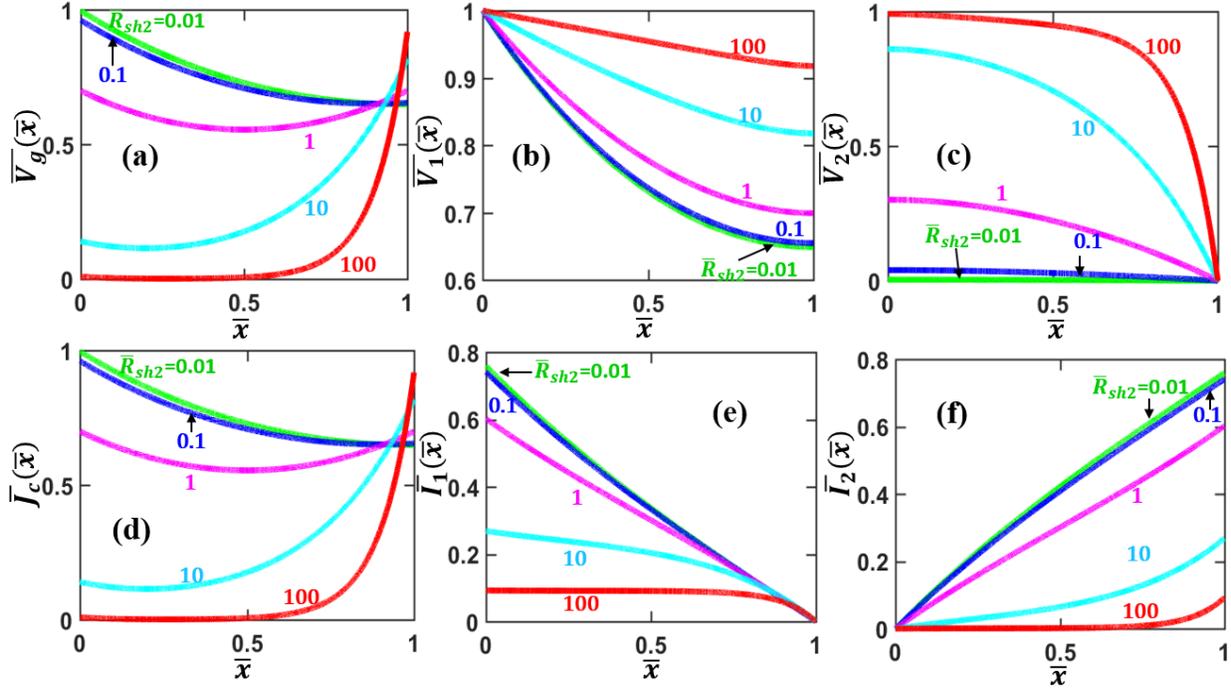

FIG. 3. Voltage and current profiles along dissimilar parallel contacts with uniform contact resistivity. (a) Voltage drop across the contact interface $\bar{V}_g(\bar{x})$, voltage along (b) contact member 1 (MN), $\bar{V}_1(\bar{x})$, and (c) contact member 2 (PQ), $\bar{V}_2(\bar{x})$, (d) current density across the contact interface $\bar{J}_c(\bar{x})$, current along (e) contact member 1, $\bar{I}_1(\bar{x})$, and (f) contact member 2, $\bar{I}_2(\bar{x})$, for different values of $\bar{R}_{sh2} = R_{sh2}/R_{sh1}$, for $\bar{\rho}_c = 1$. All the quantities are in their normalized forms defined in Sec. II.

Figure 3 shows the current and voltage distributions along the contact length and across the contact interface for various parallel contacts formed between dissimilar materials, $\bar{R}_{sh2} = R_{sh2}/R_{sh1}$, with fixed specific contact resistivity $\bar{\rho}_c = 1$. The voltage $\overline{V_{1,2}}$ and the current $\overline{I_{1,2}}$ along the two contact members show similar behaviors as those in Fig. 2. However, the voltage drop across the interface layer $\bar{V}_g(\bar{x})$ and the contact current density $\bar{J}_c(\bar{x})$ are no longer symmetric, as shown in Figs. 3(a) and 3(d), respectively. When $\bar{R}_{sh2} < 1$, the maximum of $\bar{V}_g(\bar{x})$ and $\bar{J}_c(\bar{x})$ occurs at $\bar{x} = 0$; when $\bar{R}_{sh2} > 1$, the maximum of $\bar{V}_g(\bar{x})$ and $\bar{J}_c(\bar{x})$ occurs at $\bar{x} = 1$. This current crowding effect can again be explained by the fact that current flows are self-arranged to take the least resistive path in the circuit by adjusting the current distribution according to the local resistance.

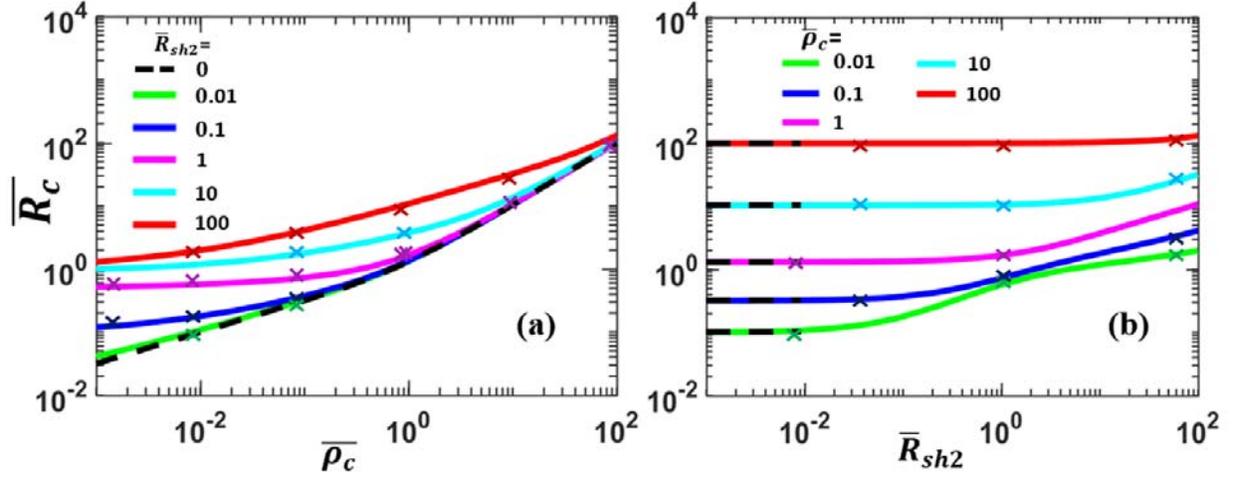

FIG. 4. Normalized contact resistance $\overline{R_c}$ of the parallel contact (Fig. 1). $\overline{R_c}$ as a function of (a) normalized specific contact resistivity, $\bar{\rho}_c$ and (b) normalized sheet resistance of contacting member 2, $\bar{R}_{sh2}$. Dashed lines are for Eq. (9), the limiting case of $\bar{R}_{sh2} \rightarrow 0$. The cross symbols are from COMSOL[45] 2D simulations. The length and height of both upper and lower contacting members are assumed to be 20 nm and 10 nm respectively, and the thickness of the resistive interfacial layer is assumed to be 0.5 nm. The resistivities of the upper and lower contact members are in the range of $10^{-9}\Omega m - 10^{-7}\Omega m$, and the resistivity of the interface layer is in the range of $10^{-9}\Omega m - 10^{-5}\Omega m$.

The normalized contact resistance, $\overline{R_c}$ calculated from Eq. (8) is plotted in Fig. 4 for various $\bar{\rho}_c$ and $\bar{R}_{sh2}$. It is clear that $\overline{R_c}$ increases with both $\bar{\rho}_c$ and $\bar{R}_{sh2}$. In general, the contact resistance $\overline{R_c}$ depends more strongly on the the specific contact resistivity of the interfacial layer $\bar{\rho}_c$ than on the sheet resistance ratio of the contact members $\bar{R}_{sh2}$. For the special case of $\bar{R}_{sh2} = 0$, Eq. (8) becomes,

$$\overline{R_c} = \frac{\coth q}{q}, \quad (9)$$

with $q = L/\lambda_0 = 1/\sqrt{\bar{\rho}_c}$, which is also plotted in Fig. 4. Note that Eq. (9) is identical to the expression typically used for metal-semiconductor contact[10,42].

To verify the results obtained from our analytical solution, we performed numerical simulations using the COMSOL multiphysics software[45], for various combinations of $\bar{R}_{sh2}$ and $\bar{\rho}_c$ on the geometry shown in Fig. 1. The finite-element-method (FEM) based COMSOL 2D simulation results are included in Fig. 4 (cross symbols), showing excellent agreement with our theory. The convergence iteration error was less than $10^{-9}$ for each point.

**Case 2: Specific resistivity $\rho_c$ varies linearly along the contact length $L$**

We assume the specific resistivity varies linearly along the contact length (Fig. 1) as $\bar{\rho}_c(\bar{x}) = 1 + A\bar{x}$. By solving Eqs. (5) and (6) numerically, we obtain the current and voltage distributions along the contact interface, as shown in Fig. 5. As $A$ increases, the overall contact interface becomes more resistive, therefore, the voltage drop $\bar{V}_g(\bar{x})$ across the interface layer increases (Fig. 5a), whereas the current density $\bar{J}_C(\bar{x})$ across the interface layer decreases in general (Fig. 5d). The maximum $\bar{V}_g$ occurs at the contact edge with the highest specific resistivity $\bar{\rho}_c$ (i.e., at $\bar{x} = 0$ when $A < 0$, and at $\bar{x} = 1$ when $A > 0$), while the maximum interface current $\bar{J}_C$ occurs at the contact edge with the lowest $\bar{\rho}_c$ (i.e., at $\bar{x} = 1$ when $A < 0$, and at $\bar{x} = 0$ when $A > 0$). The effects of $A$ on the voltage $\overline{V_{1,2}}$ and the current $\overline{I_{1,2}}$ along the two contact members are also shown in Figs. 5 (b), (c), (e) and (f), respectively.

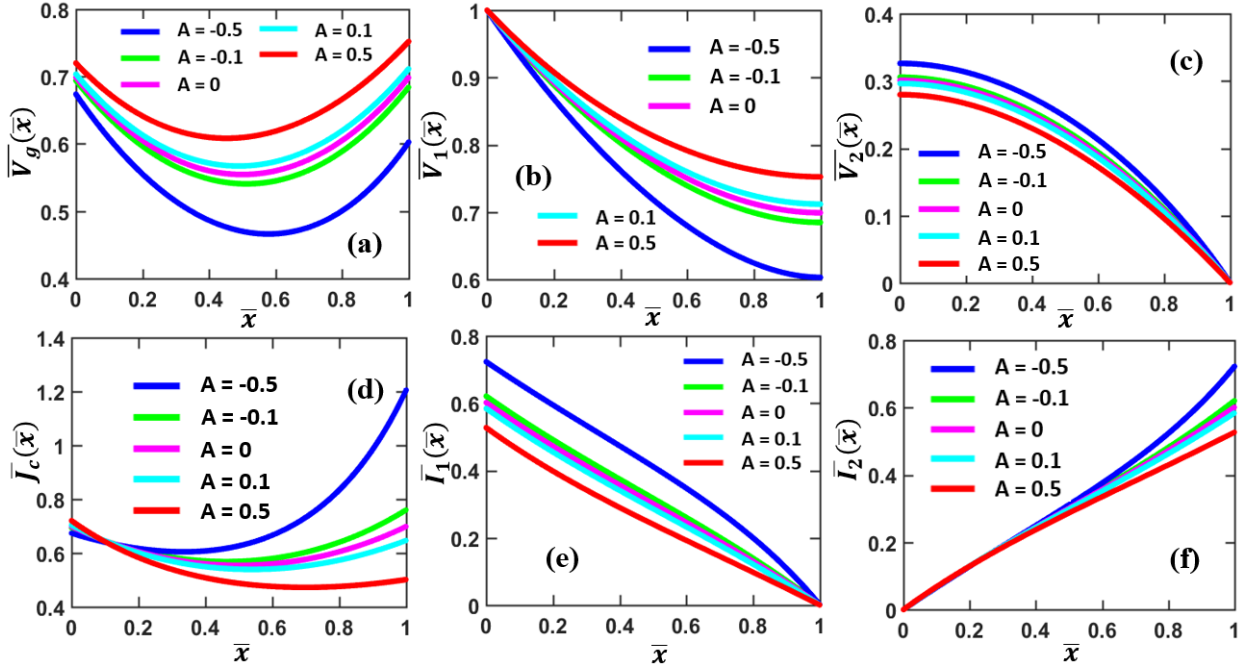

FIG. 5. Voltage and current profiles along similar parallel contacts with linearly varying contact resistivity. (a) Voltage drop across the contact interface $\bar{V}_g(\bar{x})$, voltage along (b) contact member 1 (MN), $\bar{V}_1(\bar{x})$, and (c) contact member 2 (PQ), $\bar{V}_2(\bar{x})$, (d) current density across the contact interface $\bar{J}_C(\bar{x})$, current along (e) contact member 1, $\bar{I}_1(\bar{x})$, and (f) contact member 2, $\bar{I}_2(\bar{x})$, for linear specific contact resistivity $\bar{\rho}_c(\bar{x}) = 1 + A\bar{x}$ with different linear constant $A$, for $\bar{R}_{sh2} = R_{sh2}/R_{sh1} = 1$. All the quantities are in their normalized forms defined in Sec. II.

The normalized contact resistance, $\bar{R}_c$ calculated from Eq. (4) for linear specific contact resistivity $\bar{\rho}_c(\bar{x}) = 1 + A\bar{x}$ is plotted in Fig. 6. As $A$ increases, $\bar{R}_c$ increases, since the contact interface becomes more resistive. As $\bar{R}_{sh2}$ increases, the contact resistance $\bar{R}_c$ depends more strongly on the linear constant $A$.

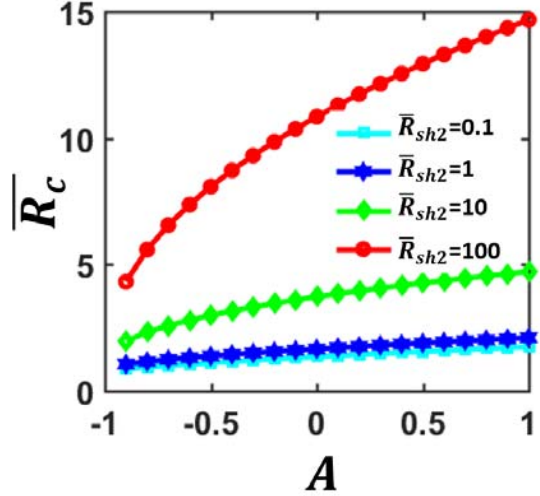

FIG. 6. Normalized contact resistance $\overline{R_c}$ of the parallel contact (Fig. 1) with linear specific contact resistivity $\bar{\rho}_c(\bar{x}) = 1 + A\bar{x}$, for various value of $\bar{R}_{sh2} = R_{sh2}/R_{sh1}$.

**Case 3: Tunneling contact resistance**

Here, we assume the parallel contacts are formed through a tunneling interface layer between the two contact members. For simplicity, we have made the following assumptions: 1) the thickness of interfacial insulating film in the contact area is uniform, and 2) the insulating film is sufficiently thin (in the nano- or subnano-meter scale) so that charge trappings are ignored [51,52].

For dissimilar contact members, the (normalized) current density at any location along the contact from contact member 1 to contact member 2 is calculated using Simmons' formula [37],

$$\bar{J}_c(\bar{x}) = B\left[\bar{\varphi}_I e^{-A\Delta\bar{y}\sqrt{\bar{\varphi}_I}} - \left(\bar{\varphi}_I + \bar{V}_g(\bar{x})\right) e^{-A\Delta\bar{y}\sqrt{\bar{\varphi}_I + \bar{V}_g(\bar{x})}}\right] \qquad (10)$$

where $\bar{V}_g(\bar{x}) = \bar{V}_1(\bar{x}) - \bar{V}_2(\bar{x})$ is the local voltage drop across the contact interface at $\bar{x}$, A = $1.025\sqrt{eV_0\,[eV]}D[\text{Å}]$, $B = 615\frac{L^2[\mu m]R_{sh1}[\Omega/\square]}{D^2[\text{Å}](\Delta\bar{y})^2}$ and $\Delta\bar{y} = \bar{y}_2 - \bar{y}_1$. Definitions of $\bar{\varphi}_I$, $\bar{y}_1$ and $\bar{y}_2$ for forward bias (when lower work function contacting member is given positive bias) are given below,

$\bar{\varphi}_I = \bar{\varphi}_2 - (\bar{V}_g(\bar{x}) + \Delta\bar{\varphi})\frac{\bar{y}_1 + \bar{y}_2}{2} - \frac{1.15\bar{\lambda}}{\bar{y}_2 - \bar{y}_1}\ln\left(\frac{\bar{y}_2(1-\bar{y}_1)}{\bar{y}_1(1-\bar{y}_2)}\right)$, where $\Delta\bar{\varphi} = \bar{\varphi}_2 - \bar{\varphi}_1$, $\bar{\varphi}_1 = \frac{\varphi_1}{eV_0}$, $\bar{\varphi}_2 = \frac{\varphi_2}{eV_0}$, $\varphi_1 = W_1 - \chi$ and $\varphi_2 = W_2 - \chi$. $W_1$ and $W_2$ are the work functions of contacting member 1 and 2 respectively, $\chi$ is the electron affinity of the insulating layer, which is 0 for vacuum. For $\bar{V}_g(\bar{x}) \leq \bar{\varphi}_1$: $\bar{y}_1 = \frac{1.2\,\bar{\lambda}}{\bar{\varphi}_2}$, $\bar{y}_2 = 1 - \frac{9.2\bar{\lambda}}{3\bar{\varphi}_2 + 4\bar{\lambda} - 2(\bar{V}_g(\bar{x}) + \Delta\bar{\varphi})} + \bar{y}_1$; and for $\bar{V}_g(\bar{x}) > \bar{\varphi}_1$: $\bar{y}_1 = \frac{1.2\,\bar{\lambda}}{\bar{\varphi}_2}$, $\bar{y}_2 = \frac{\bar{\varphi}_2 - 5.6\bar{\lambda}}{(\bar{V}_g(\bar{x}) + \Delta\bar{\varphi})}$, where $\bar{\lambda} = \frac{2.49}{\varepsilon_r D[\text{Å}]eV_0[eV]}$.

On the other hand, the definitions of $\bar{\varphi}_I$, $\bar{y}_1$ and $\bar{y}_2$ for reverse bias (when higher work function contacting member is given positive bias) are: $\bar{\varphi}_I = \bar{\varphi}_1 + (\Delta\bar{\varphi} - \bar{V}_g(\bar{x}))\frac{\bar{y}_1+\bar{y}_2}{2} - \frac{1.15\bar{\lambda}}{\bar{y}_2-\bar{y}_1}\ln\left(\frac{\bar{y}_2(1-\bar{y}_1)}{\bar{y}_1(1-\bar{y}_2)}\right)$, for $0 < \bar{V}_g(\bar{x}) \le \Delta\bar{\varphi}$: $\bar{y}_1 = \frac{9.2\bar{\lambda}}{3\bar{\varphi}_1+4\bar{\lambda}-(\bar{V}_g(\bar{x})-\Delta\bar{\varphi})} - \frac{1.2\,\bar{\lambda}}{\bar{\varphi}_2-\bar{V}_g(\bar{x})}$, $\bar{y}_2 = 1 - \frac{1.2\,\bar{\lambda}}{\bar{\varphi}_2-\bar{V}_g(\bar{x})}$; for $\Delta\bar{\varphi} < \bar{V}_g(\bar{x}) \le \bar{\varphi}_2$: $\bar{y}_1 = \frac{1.2\,\bar{\lambda}}{\bar{\varphi}_1}$, $\bar{y}_2 = 1 - \frac{9.2\bar{\lambda}}{3\bar{\varphi}_1+4\bar{\lambda}-2(\bar{V}_g(\bar{x})-\Delta\bar{\varphi})} + \bar{y}_1$; and for $\bar{V}_g(\bar{x}) > \bar{\varphi}_2$: $\bar{y}_1 = \frac{1.2\,\bar{\lambda}}{\bar{\varphi}_1}$, $\bar{y}_2 = \frac{\bar{\varphi}_1-5.6\bar{\lambda}}{(\bar{V}_g(\bar{x})-\Delta\bar{\varphi})}$.

For the special case of the same material for contact members 1 and 2 [36], in Eq. (10), $\bar{\varphi}_I = \bar{\varphi}_0 - \bar{V}_g(\bar{x})\frac{\bar{y}_1+\bar{y}_2}{2} - \frac{1.15\bar{\lambda}}{\bar{y}_2-\bar{y}_1}\ln\left(\frac{\bar{y}_2(1-\bar{y}_1)}{\bar{y}_1(1-\bar{y}_2)}\right)$ where $\bar{\varphi}_0 = \frac{\varphi_0}{eV_0}$, $\varphi_0 = W - \chi$, $W$ is the work function of contacting member 1 and 2, and, $\bar{y}_1 = \frac{1.2\,\bar{\lambda}}{\bar{\varphi}_0}$, $\bar{y}_2 = 1 - \frac{9.2\bar{\lambda}}{3\bar{\varphi}_0+4\bar{\lambda}-2\bar{V}_g(\bar{x})} + \bar{y}_1$ for $\bar{V}_g(\bar{x}) \le \bar{\varphi}_0$, $\bar{y}_2 = \frac{\bar{\varphi}_0-5.6\bar{\lambda}}{\bar{V}_g(\bar{x})}$ for $\bar{V}_g(\bar{x}) \le \bar{\varphi}_0$. Note that we use Simmon's formula, Eq. (10) here for simplicity, which is reliable only when the barrier height is relative high and the gap voltage is low in the direct tunneling regime [11]. More accurate results for the tunneling current may be calculated using quantum models by solving the coupled Schrodinger equation and Poisson equation with the inclusion of space charge and exchange-correlation effects [11,25].

We keep the normalization consistent with our previous calculations in Sec. II. For a given parallel tunneling contact (Fig. 1), the inputs of our model are the applied voltage $V_0$, sheet resistance ($R_{sh1}, R_{sh2}$) and work function ($W_1, W_2$) of contacting members 1 and 2, permittivity ($\varepsilon_r$), thickness ($D$), and electron affinity ($\chi$) of the interfacial insulator layer, and the contact length $L$. Using Eq. (10), the specific contact resistivity is obtained from $\bar{\rho}_c(\bar{x}) = \bar{V}_g(\bar{x})/\bar{J}_c(\bar{x})$, which is inserted into the TLM equations, Eqs. (5) and (6), to give a self-consistent solution to the voltage and current profiles, as well as the contact resistance for the parallel tunneling contact.

We consider CNT-vacuum-CNT parallel contact as an example. Both contact members are made of the same single-walled CNTs. Using the typical value of linear resistivity of single-walled CNT $\rho_L = 20$ k$\Omega$/μm [53,54], and diameter (or the width $w$) of 3 nm, an equivalent sheet resistance for both CNT contact members are estimated as $R_{sh1} = R_{sh2} = \rho_L w = 60$ $\Omega/\square$, where the unit of the sheet resistance $\Omega/\square$ means "ohm per square" [10,42,49]. The work function of CNTs is $W_1 = W_2 = 4.5$ eV [55]. The interfacial layer is assumed to be vacuum (relative permittivity $\varepsilon_r = 1.0$, and electron affinity $\chi = 0$). The voltage drop $V_g(x)$ across and the tunneling current density $J_c(x)$ through the contact interface are shown in Fig. 7 for various contact length $L$, vacuum gap distance $D$, and applied voltage $V_0$. The profiles of both $V_g(x)$ and $J_c(x)$ are symmetric about the center of the contact, as expected for similar contact members (similar to Figs. 2a and 2d above). As the contact length $L$ increases, the local voltage drop $V_g(x)$ across the contact interface decreases, so does the tunneling current density $J_c(x)$, as shown in Figs. 7a and 7b. However, the total current in the contact structure, $I_{tot} = \int_0^L J_c(x)dx$ increases with $L$, since the total contact resistance of the tunneling junction decreases as the contact length increases (cf. Fig. 8a below). As shown in Figs. 7c and 7d, when the gap distance $D$ increases, the voltage drop $V_g(x)$ increases, but the current density $J_c(x)$ decreases, which is because the tunneling junction becomes more resistive [11,36]. Figures 7e and 7f shows both voltage drop $V_g(x)$ and current density $J_c(x)$ increase when the applied voltage $V_o$ increases. More importantly, both $V_g(x)$ and $J_c(x)$ exhibit a stronger spatial dependence as $V_0$ increases. This strong voltage dependence of electrical properties of the tunneling junction is in sharp contrast with those of ohmic contacts, where the profiles of $V_g(x)$ and $J_c(x)$, and the total contact resistance is independent of the applied voltage, and the current density scales linearly with the voltage drops, as discussed in Cases 1 and 2 above.

Also plotted in Fig. 7 are the analytical results from Eq. (7), by assuming constant tunneling contact resistivity across the contact length $L$ (i.e. the typically assumed one-dimensional tunneling junction [24]), by (a), setting $V_g = V_0$ and using Eq. (10) (dashed lines) and (b), using ohmic approximations for the tunneling junction, in the limit of $V_g \to 0$ (dotted lines)[36,37]. In the latter case, the tunneling current density is a linear function of $V_g$.

$$\bar{J}_c(\bar{x}) = B\bar{\varphi}_I e^{-A\Delta\bar{y}\sqrt{\bar{\varphi}_I}} \bar{V}_g(\bar{x}) \, , V_g \to 0 \quad (11),$$

where $= 315.60 \sqrt{V_0 \frac{L^2[\mu m] R_{sh1}[\frac{\Omega}{\Box}]}{D[\text{Å}]\Delta y}}$. $A$ and $\Delta\bar{y}$ are the same as for Eq (10). $\bar{\varphi}_I$ is calculated from the same expression for Eq (10) by setting $V_g = 0$. $\bar{y}_1 = \frac{1.2\,\bar{\lambda}}{\bar{\varphi}_2}$, $\bar{y}_2 = 1 - \frac{9.2\bar{\lambda}}{3\bar{\varphi}_2 + 4\bar{\lambda} - 2\Delta\bar{\varphi}} + \bar{y}_1$ for forward bias; $\bar{y}_1 = \frac{9.2\bar{\lambda}}{3\bar{\varphi}_1 + 4\bar{\lambda} + \Delta\bar{\varphi}} - \frac{1.2\,\bar{\lambda}}{\bar{\varphi}_2}$, $\bar{y}_2 = 1 - \frac{1.2\,\bar{\lambda}}{\bar{\varphi}_2}$ for reverse bias; and $\bar{y}_1 = \frac{1.2\,\bar{\lambda}}{\bar{\varphi}_0}$, $\bar{y}_2 = 1 - \frac{1.2\,\bar{\lambda}}{\bar{\varphi}_0}$ for similar contacting members. $\bar{\lambda}$, $\Delta\bar{\varphi}$, $\bar{\varphi}_2$, $\bar{\varphi}_1$, $\bar{\varphi}_0$ have the same definition as in Eq. (10). It is found that both assumptions of constant contact resistivity are not sufficiently reliable, especially when the tunneling thickness $D$ decreases or the applied voltage $V_o$ increases. As the tunneling junction resistance becomes nonlinear in these cases, it is necessary to use the coupled TLM equations, Eqs. (5) and (6), and the localized tunneling equation, Eq. (10), to provide more accurate predictions.

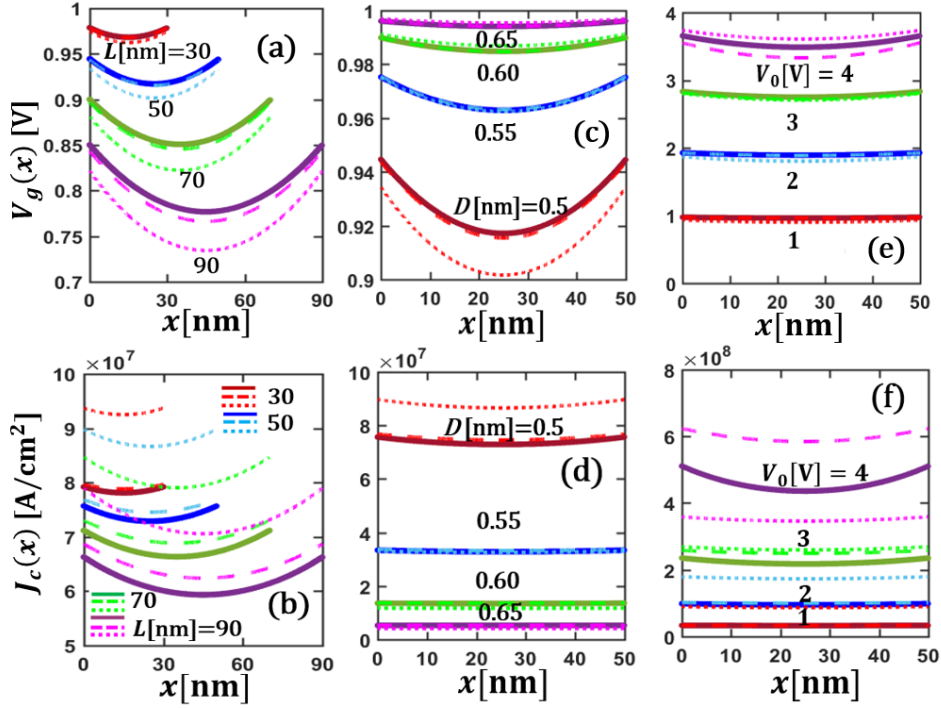

FIG. 7. Similar material CNT-vacuum-CNT parallel tunneling contacts. (a) Voltage drop across the contact interface $V_g(x)$, and (b) tunneling current density across the contact interface $J_c(x)$ for different contact length $L$, with fixed $V_0 = 1$V, and $D = 0.5$ nm; (c) $V_g(x)$ and (d) $J_c(x)$ for different $D$, with fixed $V_0 = 1$V and $L = 50$ nm; (e) $V_g(x)$ and (f) $J_c(x)$ for different applied voltage $V_0$ with fixed $D = 0.55$ nm, and $L = 50$ nm. All the material properties are specified in the main text. Solid lines are for self-consistent numerical calculations using Eqs. 5, 6, and 10, dashed and dotted lines are for analytical calculations from Eq. 7 with $\rho_c$ calculated using $V_g = V_0$ in Eq 10 and ohmic approximations for the tunneling junction, Eq. 11, in the limit of $V_g \to 0$, respectively.

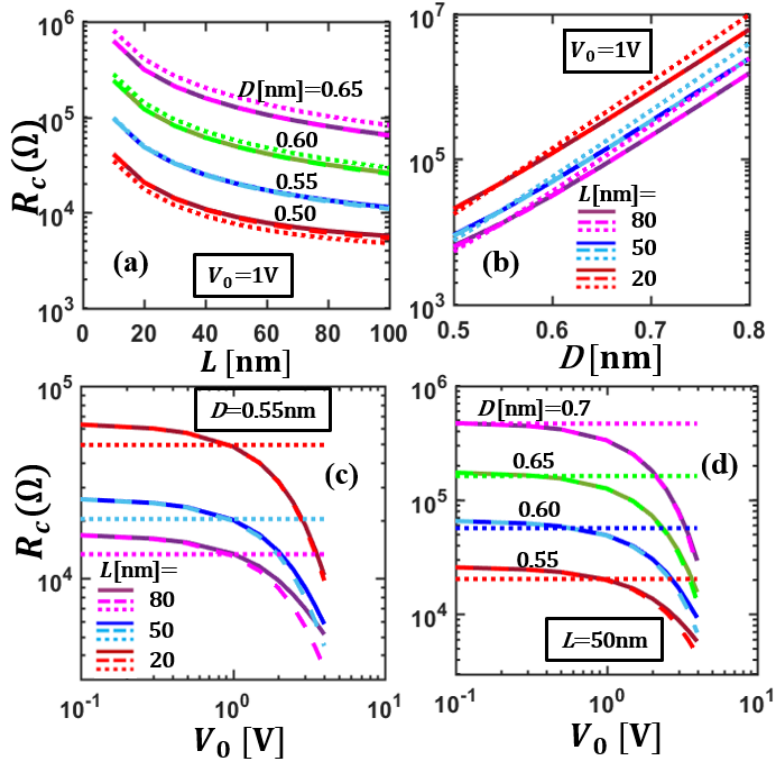

FIG. 8. The total contact resistance $R_c$ of the CNT-vacuum-CNT parallel contact. Contact resistance is plotted as a function of (a) contact length, $L$, for different insulating layer thickness, $D$, (b) $D$, for different $L$, for a fixed applied voltage, $V_0 = 1V$ ; (c) and (d) applied voltage $V_0$ for different $L$ and $D$ respectively, in CNT-vacuum-CNT contacts. Solid lines are for self-consistent numerical calculations using Eqs. 5, 6, and 10, dashed and dotted lines are for analytical calculations from Eq. 8 with $\rho_c$ calculated using $V_g = V_0$ in Eq 10 and ohmic approximations for the tunneling junction, Eq. 11, in the limit of $V_g \to 0$, respectively.

The total contact resistance $R_c$ of the CNT-vacuum-CNT parallel contact is shown in Fig. 8, as functions of contact length $L$, vacuum gap distance $D$, and applied voltage $V_0$. The total contact resistance $R_c$ increases very rapidly with increasing insulating layer thickness, $D$, and decreases with contact length, $L$. For the low applied voltage regime ($V_0 < 0.3$ V), $R_c$ is almost independent of $V_0$, as shown in Figs. 8c and 8d. When the applied voltage $V_0 > 0.3$ V, $R_c$ decreases sharply with $V_0$. This is because the junction is no longer ohmic and the tunneling resistivity $\rho_c$ decreases nonlinearly with the junction voltage, as a function of position along the contact length. Ohmic approximations (Eqs. 8, 11) fail to give accurate results in the latter case and it is necessary to use the self-consistent numerical model. As $L$ increases, the dependence of contact resistance on $L$ becomes less significant. Similar profiles of contact resistance with $L$ were observed in other experimental and theoretical works [10,22,31,42,44]. The contact resistance lies between 5 kΩ to 10 MΩ for the cases shown in Fig. 8, which agrees with previously reported experimental and theoretical works [24,26,29]. The existing 1D models give an inaccurate estimation of the contact resistance because they do not consider the variation of tunneling current density along the contact length.

Next, we extend our calculations for contacts of CNT with different metals – calcium (Ca), aluminum (Al), copper (Cu) and gold (Au). The work functions of Ca, Al, Cu and Au are taken as 2.9, 4.08, 4.7 and 5.1 eV respectively [56]. The work functions and dimensions of the CNT are kept same as before. In addition, the dimensions of the CNT and contacting-metal-2 are assumed to be same (width of 3 nm, thickness of 3 nm) for the simplicity of calculations. The resistivity of Ca, Al, Cu and Au are known to be $3.36 \times 10^{-8}\ \Omega m$, $2.7 \times 10^{-8}\ \Omega m$, $1.68 \times 10^{-8}\ \Omega m$ and $2.2 \times 10^{-8}\ \Omega m$ respectively [56,57].

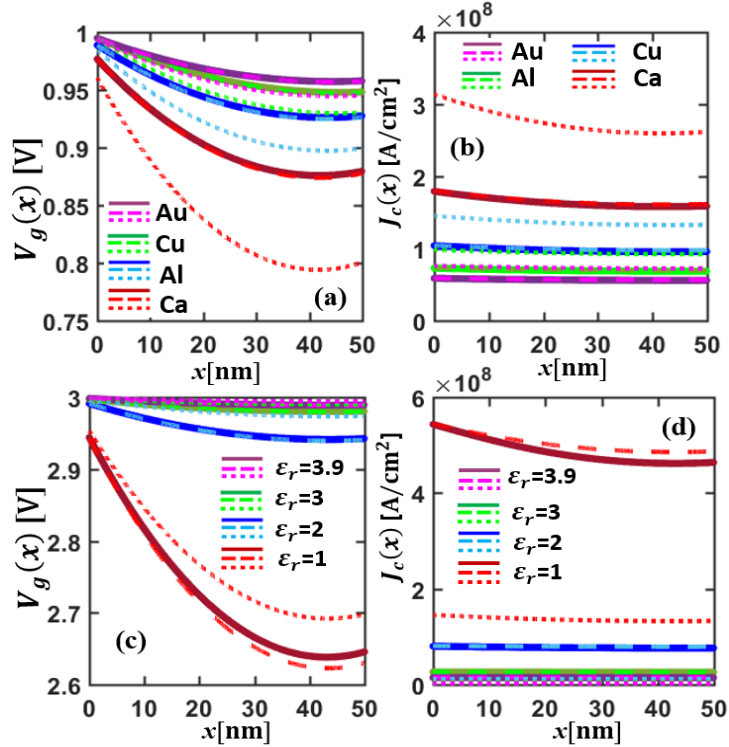

FIG. 9. Dissimilar material CNT-insulator-metal parallel tunneling contacts. (a) Voltage drop across the interfacial insulating layer $V_g(x)$, and (b) tunneling current density $J_c(x)$, in CNT-insulator-Metal contacts, for fixed $D = 0.5$nm, $L = 50$nm, $V_0 = 1$V and different contacting metals (Ca, Al, Cu, Au). (c) $V_g(x)$, and (d) $J_c(x)$, in CNT-insulator-Al contacts, for different insulating layer permittivity $\varepsilon_r$, with fixed $D = 0.5$nm, $L = 50$nm, $V_0 = 3$V. Solid lines are for self-consistent numerical calculations using Eqs. 5, 6, and 10, dashed and dotted lines are for analytical calculations from Eq. 7 with $\rho_c$ calculated using $V_g = V_0$ in Eq 10 and ohmic approximations for the tunneling junction, Eq. 11, in the limit of $V_g \rightarrow 0$, respectively.

Figure 9 shows the effects of the work function of contacting member 2 ($W_2$) and the permittivity of the thin insulating layer ($\varepsilon_r$), on the current and voltage characteristics in CNT-insulator-metal contacts. As the two contact members are different, the voltage drop $V_g(x)$ and the tunneling current density $J_c(x)$ are no longer symmetric along the contact length $L$. Figure 9(a) and 9(b)
16

show that the voltage drop increases and the tunneling current density decreases with increasing $W_2$. Figure 9(c) and 9(d) show that the voltage drop increases and the tunneling current density reduces significantly when the permittivity of the insulating layer increases from 1 to 3.9. Analytical solutions obtained by assuming constant tunneling resistivity along the contact length are also included, similar to the previous cases of Fig. 7. In general, for the chosen value of $D = 0.5$ nm, the ohmic approximations using Eq. 11 do not yield accurate results. The constant tunneling resistivity approximation using Eq. 10 by setting $V_g = V_0$ could be a good approximation for the self-consistent TLM model (Eqs. 5, 6, and 10), for tunneling layers with higher permittivity $\varepsilon_r$.

Figure 10 shows the contact resistance (in Ω) for various contact metals and tunneling films for CNT-insulator-metal contacts. Contact resistance increases with insulating layer thickness $D$, insulating layer permittivity $\varepsilon_r$ and work function of contacting member $W_2$. It decreases with contact length $L$, as in the similar contacts in Fig. 8. The potential barrier in the insulating layer increases with the increase of work function of the contact metal, resulting in lower tunneling current and higher contact resistance.

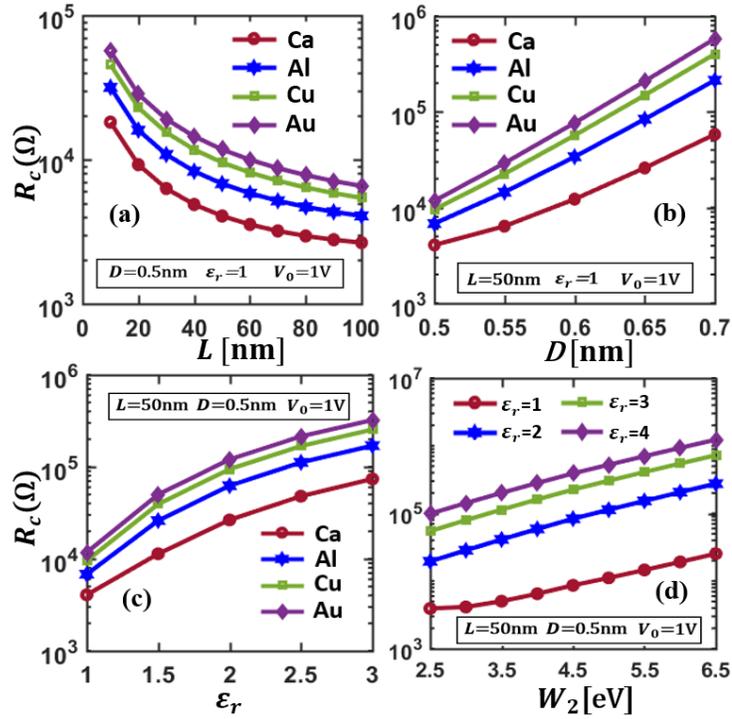

FIG. 10. The total contact resistance $R_c$ of the CNT-insulator-metal parallel contact. Contact resistance is plotted as a function of (a) contact length $L$, (b) insulator layer thickness $D$ and (c) insulator layer permittivity $\varepsilon_r$, for CNT-insulator-metal contacts for different contacting metals (Ca, Al, Cu, Au). (d) Contact resistance as a function of work function of contacting member 2 ($W_2$). The material properties and dimensions for (a)-(c) are specified in the text (the same as in Fig. 9). For (d), the resistivity of contacting member 2 is assumed to be $2.0 \times 10^{-8}$ $\Omega m$. The results are from the self-consistent numerical calculations using Eqs. 5, 6, and 10.

## IV. SUMMARY

In this paper, we proposed a self-consistent model to characterize partially overlapped parallel contacts. Our model considers the spatial variation of contact resistivity along the contact structure. We solved the TLM equations for three cases: 1) constant specific contact resistivity, 2) linearly varying specific contact resistivity, and 3) spatial dependent specific contact resistivity along the contact length due to current tunneling. The analytical solutions for the first case have been verified with FEM based numerical simulations. Our study provides a thorough understanding of the contact tunneling resistance, current and voltage distributions across nano and sub-nano scale MIM junctions in parallel electrical contacts. The effects of contact geometry (i.e. dimension of the contact, and distance between the contact electrodes), and material properties (i.e. work function, sheet resistance of the contact members, and permittivity of the insulating layer) on the spatial distributions of currents and voltages across these contacts, and the overall contact resistance are studied in detail. While predominately classical in nature, the inclusion of tunneling current starts to address quantum effects in these small scale objects.

It is found that in general the ohmic approximation of tunneling junctions (Eq. 11) is not reliable for predicting the contact resistance of parallel tunneling contacts. The one-dimensional (1D) tunneling junction models (Eq. 10 with constant voltage across the whole junction) are good approximations of the parallel contacts only when the thickness $D$ or the permittivity $\varepsilon_r$ of the tunneling film is relatively large, or the applied voltage across the contact $V_0$ is relatively small. When the 1D models become unreliable for small $D$ or $\varepsilon_r$, or large $V_0$, the self-consistent TLM equations coupled with the tunneling current (Eqs. 5, 6 and 10) need to be used to accurately characterize the parallel tunneling contacts.

The parallel tunneling contact in this work may be considered as the basic building block to better understand the macroscopic electrical conductivity of CNT fibers, which contains a very large number of such parallel contacts between individual CNTs. Furthermore, our study elucidates key parameters for parallel electrical contacts over a wide range of spatially dependent contact resistivity, which paves the way to strategically design of contact structures with controlled current distribution profiles and contact resistance, by spatially varying the contact layer properties and geometry.

In this formulation, we have ignored the effects of space charge and exchange-correlation inside the tunneling gap [11,40]. We have also ignored possible charge trapping inside contact junctions. The model is assumed two-dimensional, where the effects of the transverse dimension are neglected. These issues will be the subjects of future studies. It is important to note that the transmission line model (TLM) is only a simplified approximation of the 2D electrical contacts, where the current crowding and the fringing fields near the contact corners cannot be fully accounted for [10,49]. In order to accurately evaluate these effects as well as the impact of finite thickness in the contact members and the contact junction, and the possible parallel component of current flows in the interface layer, field solution methods need to be used [10,49,50,58]. Future studies may also include the effects of various contact geometry, insulator layer non-uniformities and AC response on the electrical properties of tunneling type contacts. An additional feature might include the role of capacitance and inductance in nano- and micro-scale structures, especially when a large contact resistance when coupled with these reactive effects might introduce new time scales into time-dependent dynamic problems.

## METHODS

N. A.

## ACKNOWLEDGMENTS

The work is supported by the Air Force Office of Scientific Research (AFOSR) YIP Award No. FA9550-18-1-0061.

## AUTHOR CONTRIBUTIONS

Zhang and Luginsland conceived the problem. Zhang and Banerjee did the theoretical formulation. Banerjee did the numerical computation and data collection. All authors jointly wrote the paper.

## ADDITIONAL INFORMATION

Competing interests: The authors declare no competing interests

15. Behabtu, N. *et al.* Strong, Light, Multifunctional Fibers of Carbon Nanotubes with Ultrahigh Conductivity. *Science* **339**, 182–186 (2013).

16. Li, Q. W. *et al.* Structure-Dependent Electrical Properties of Carbon Nanotube Fibers. *Advanced Materials* **19**, 3358–3363 (2007).

17. Nieuwoudt, A. & Massoud, Y. On the Optimal Design, Performance, and Reliability of Future Carbon Nanotube-Based Interconnect Solutions. *IEEE Transactions on Electron Devices* **55**, 2097–2110 (2008).

18. Tang, J. *et al.* Flexible CMOS integrated circuits based on carbon nanotubes with sub-10 ns stage delays. *Nature Electronics* **1**, 191–196 (2018).

19. Peng, L.-M. A new stage for flexible nanotube devices. *Nature Electronics* **1**, 158–159 (2018).

20. He, X. *et al.* Wafer-scale monodomain films of spontaneously aligned single-walled carbon nanotubes. *Nat Nanotechnol* **11**, 633–638 (2016).

21. Zaumseil, J. Single-walled carbon nanotube networks for flexible and printed electronics. *Semicond. Sci. Technol.* **30**, 074001 (2015).

22. Buldum, A. & Lu, J. P. Contact resistance between carbon nanotubes. *Physical Review B* **63**, (2001).

23. Kilbride, B. E. *et al.* Experimental observation of scaling laws for alternating current and direct current conductivity in polymer-carbon nanotube composite thin films. *Journal of Applied Physics* **92**, 4024–4030 (2002).

24. Li, C., Thostenson, E. T. & Chou, T.-W. Dominant role of tunneling resistance in the electrical conductivity of carbon nanotube–based composites. *Appl. Phys. Lett.* **91**, 223114 (2007).

FIG. 1. A parallel, partially overlapped electric contact. The contacts are formed between (a) nanotube or nanowire 1 and 2, and (b) thin film 1 and 2; (c) side view of the contact; (d) its transmission line model. In (a), (b) and (c) a thin resistive interface layer (or a tunneling layer of permittivity $\varepsilon_r$) is sandwiched between the two contacting members.

FIG. 2. Voltage and current profiles along similar parallel contacts with uniform contact resistivity. (a) Voltage drop across the contact interface $\bar{V}_g(\bar{x})$, voltage along (b) contact member 1 (MN), $\bar{V}_1(\bar{x})$, (c) contact member 2 (PQ), $\bar{V}_2(\bar{x})$, (d) current density across the contact interface $\bar{J}_c(\bar{x})$, current along (e) contact member 1, $\bar{I}_1(\bar{x})$, and (f) contact member 2, $\bar{I}_2(\bar{x})$, for different values of specific contact resistivity $\bar{\rho}_c$, for $\bar{R}_{sh2} = R_{sh2}/R_{sh1} = 1$. All the quantities are in their normalized forms defined in Sec. II.

FIG. 3. Voltage and current profiles along dissimilar parallel contacts with uniform contact resistivity. (a) Voltage drop across the contact interface $\bar{V}_g(\bar{x})$, voltage along (b) contact member 1 (MN), $\bar{V}_1(\bar{x})$, and (c) contact member 2 (PQ), $\bar{V}_2(\bar{x})$, (d) current density across the contact interface $\bar{J}_c(\bar{x})$, current along (e) contact member 1, $\bar{I}_1(\bar{x})$, and (f) contact member 2, $\bar{I}_2(\bar{x})$, for different values of $\bar{R}_{sh2} = R_{sh2}/R_{sh1}$, for $\bar{\rho}_c = 1$. All the quantities are in their normalized forms defined in Sec. II.

FIG. 4. Normalized contact resistance $\bar{R}_c$ of the parallel contact (Fig. 1). $\bar{R}_c$ as a function of (a) normalized specific contact resistivity, $\bar{\rho}_c$ and (b) normalized sheet resistance of contacting member 2, $\bar{R}_{sh2}$. Dashed lines are for Eq. (9), the limiting case of $\bar{R}_{sh2} \to 0$. The cross symbols are from COMSOL[45] 2D simulations. The length and height of both upper and lower contacting members are assumed to be 20 nm and 10 nm respectively, and the thickness of the resistive interfacial layer is assumed to be 0.5 nm. The resistivities of the upper and lower contact members are in the range of $10^{-9} \Omega\text{m} - 10^{-7} \Omega\text{m}$, and the resistivity of the interface layer is in the range of $10^{-9} \Omega\text{m} - 10^{-5} \Omega\text{m}$.

FIG. 5. Voltage and current profiles along similar parallel contacts with linearly varying contact resistivity. (a) Voltage drop across the contact interface $\bar{V}_g(\bar{x})$, voltage along (b) contact member 1 (MN), $\bar{V}_1(\bar{x})$, and (c) contact member 2 (PQ), $\bar{V}_2(\bar{x})$, (d) current density across the contact interface $\bar{J}_c(\bar{x})$, current along (e) contact member 1, $\bar{I}_1(\bar{x})$, and (f) contact member 2, $\bar{I}_2(\bar{x})$, for linear specific contact resistivity $\bar{\rho}_c(\bar{x}) = 1 + A\bar{x}$ with different linear constant $A$, for $\bar{R}_{sh2} = R_{sh2}/R_{sh1} = 1$. All the quantities are in their normalized forms defined in Sec. II.

FIG. 6. Normalized contact resistance $\overline{R_c}$ of the parallel contact (Fig. 1) with linear specific contact resistivity $\bar{\rho}_c(\bar{x}) = 1 + A\bar{x}$, for various value of $\bar{R}_{sh2} = R_{sh2}/R_{sh1}$.

FIG. 7. Similar material CNT-vacuum-CNT parallel tunneling contacts. (a) Voltage drop across the contact interface $V_g(x)$, and (b) tunneling current density across the contact interface $J_c(x)$ for different contact length $L$, with fixed $V_0 = 1V$, and $D = 0.5$ nm; (c) $V_g(x)$ and (d) $J_c(x)$ for different $D$, with fixed $V_0 = 1V$ and $L = 50$ nm; (e) $V_g(x)$ and (f) $J_c(x)$ for different applied voltage $V_0$ with fixed $D = 0.55$ nm, and $L = 50$ nm. All the material properties are specified in the main text. Solid lines are for self-consistent numerical calculations using Eqs. 5, 6, and 10, dashed and dotted lines are for analytical calculations from Eq. 7 with $\rho_c$ calculated using $V_g = V_0$ in Eq 10 and ohmic approximations for the tunneling junction, Eq. 11, in the limit of $V_g \to 0$, respectively.

FIG. 8. The total contact resistance $R_c$ of the CNT-vacuum-CNT parallel contact. Contact resistance is plotted as a function of (a) contact length, $L$, for different insulating layer thickness, $D$, (b) $D$, for different $L$, for a fixed applied voltage, $V_0 = 1V$ ; (c) and (d) applied voltage $V_0$ for different $L$ and $D$ respectively, in CNT-vacuum-CNT contacts. Solid lines are for self-consistent numerical calculations using Eqs. 5, 6, and 10, dashed and dotted lines are for analytical calculations from Eq. 8 with $\rho_c$ calculated using $V_g = V_0$ in Eq 10 and ohmic approximations for the tunneling junction, Eq. 11, in the limit of $V_g \to 0$, respectively.

FIG. 9. Dissimilar material CNT-insulator-metal parallel tunneling contacts. (a) Voltage drop across the interfacial insulating layer $V_g(x)$, and (b) tunneling current density $J_c(x)$, in CNT-insulator-Metal contacts, for fixed $D = 0.5$nm, $L = 50$nm, $V_0 = 1V$ and different contacting metals (Ca, Al, Cu, Au). (c) $V_g(x)$, and (d) $J_c(x)$, in CNT-insulator-Al contacts, for different insulating layer permittivity $\varepsilon_r$, with fixed $D = 0.5$nm, $L = 50$nm, $V_0 = 3V$ . Solid lines are for self-consistent numerical calculations using Eqs. 5, 6, and 10, dashed and dotted lines are for analytical calculations from Eq. 7 with $\rho_c$ calculated using $V_g = V_0$ in Eq 10 and ohmic approximations for the tunneling junction, Eq. 11, in the limit of $V_g \to 0$, respectively.

FIG. 10. The total contact resistance $R_c$ of the CNT-insulator-metal parallel contact. Contact resistance is plotted as a function of (a) contact length $L$, (b) insulator layer thickness $D$ and (c) insulator layer permittivity $\varepsilon_r$, for CNT-insulator-metal contacts for different contacting metals (Ca, Al, Cu, Au). (d) Contact resistance as a function of work function of contacting member 2 ($W_2$). The material properties and dimensions for (a)-(c) are specified in the text (the same as in Fig. 9). For (d), the resistivity of contacting member 2 is assumed to be $2.0 \times 10^{-8}$ $\Omega m$. The results are from the self-consistent numerical calculations using Eqs. 5, 6, and 10.